\documentclass[12pt]{article}
\usepackage{epsfig}

\input epsf.tex
\newcommand{\be}{\begin{equation}}
\newcommand{\ee}{\end{equation}}
\newcommand{\bea}{\begin{eqnarray}}
\newcommand{\eea}{\end{eqnarray}}

\newcommand{\beq}{\begin{equation}}
\newcommand{\eeq}{\end{equation}}
\newcommand{\beqa}{\begin{eqnarray}}
\newcommand{\eeqa}{\end{eqnarray}}

\begin{document}

\bigskip 
\begin{titlepage}

\begin{flushright}
UUITP-15/02\\ 
hep-th/0211006
\end{flushright}

\vspace{1cm}

\begin{center}
{\Large\bf Can MAP and Planck map Planck physics?\\}

\end{center}
\vspace{3mm}

\begin{center}

{\large Lars Bergstr\"{o}m,$^{\scriptstyle 1}$
and
Ulf H.\ Danielsson,$^{\scriptstyle 2}$} \\

\vspace{5mm}

$^{1}$ Department of Physics, Stockholm University, 
AlbaNova University Center\\ S-106 91 
Stockholm, Sweden

\vspace{3mm}

$^{2}$ Institutionen f\"or Teoretisk Fysik, Box 803, SE-751 08
Uppsala, Sweden

\vspace{5mm}

{\tt
lbe@physto.se, ulf@teorfys.uu.se \\
}

\end{center}
\vspace{5mm}

\begin{center}
{\large \bf Abstract}
\end{center}
\noindent
We investigate whether a recently proposed modulation 
of the power spectrum of primordial density fluctuations generated
through transplankian (maybe stringy) effects during inflation can
be observed. We briefly review the mechanism leading to 
the modulation and apply it to a generic slow-roll scenario of inflation. 
We then investigate how these primordial modulation effects leave an
imprint in the cosmic microwave background radiation. Our conclusions
are that for favourable parameter values already the presently flying 
MAP satellite will have a chance
to detect such transplanckian oscillations in the pattern of temperature 
fluctuations on the sky, and that the upcoming Planck satellite will 
either detect them or put stringent limits related to the mass scale
where the new effects appear.


\vfill
\begin{flushleft}
October 2002
\end{flushleft}
\end{titlepage}
\newpage


\section{Introduction}

\bigskip

It is an intriguing possibility that effects of physics beyond the Planck
scale might be visible on cosmological scales in the spectrum of the cosmic
microwave background (CMBR) fluctuations. The physical mechanism that could
make this possible is inflation. Inflation magnifies microscopic quantum
fluctuations into cosmic size, and thereby provides seeds for structure
formation. The details of physics at the highest energy scales is therefore
reflected in the distribution of galaxies and other structures on large
scales.

In the standard inflationary scenario, initial conditions for the inflaton
field are imposed in the infinite past when the modes are infinitely small.
In this limit the effect of the inflationary horizon and the expansion of
the universe can be ignored, the space time is essentially Minkowski, and
there is a unique vacuum for the inflaton field. We will refer to this
vacuum as the Bunch-Davies vacuum. The problem with this reasoning is that
we are assuming that intuition based on field theory is always applicable
and that nothing dramatic happens at the smallest length scales, when
stringy or transplankian physics is likely to occur. Clearly this is an
assumption that needs to be carefully scrutinized. Could there be effects of
new physics that will change the predictions of inflation? In particular one
could worry about changes in the predictions of the CMBR fluctuations.
Several groups have investigated various ways of modifying high energy
physics in order to look for such modifications, [1-33].

Without a good understanding of physics near the Planck, or string, scale,
one can also take a purely empiricist point of view, as advocated in \cite
{Danielsson:2002kx}, and simply encode the new physics in the choice of
initial conditions when a mode of the inflaton has a wavelength comparable
to the scale where new physics is expected. (See also \cite
{Starobinsky:2001kn}.) Since the mode will not be infinitely small compared
to the inflationary scale, the time dependence of the background will be
important and the choice of initial conditions not unique. The effects on
physics at lower energies will then be encoded in the choice of a vacuum not
necessarily the same as the Bunch-Davies vacuum.

As will be reviewed in Section~\ref{sec:review} the general expectation is
that the effects will be of linear order in $H/\Lambda $, where $H$ is the
Hubble constant during inflation and $\Lambda $ the scale where new physics
appears. Smaller corrections would suggest some kind of fine tuning and
require knowledge of planckian physics in order to be justified. The
Bunch-Davies vacuum is still a vacuum candidate but, contrary to the case
where the initial conditions are imposed in the infinite past, it is not the
unique choice that one can argue for.

There has been extensive discussions of these results in the literature. It
has been known since a long time that there is a family of vacua in de
Sitter space that respects all the symmetries, the $\alpha $-vacua, \cite
{chernikov}\cite{Mottola:ar}\cite{Floreanini:1986tq}\cite{Bousso:2001mw}\cite
{Spradlin:2001nb}. As pointed out in \cite{Danielsson:2002qh} the initial
condition approach to the transplanckian problem allows for a discussion of
many of the transplanckian effects in terms of these vacua. In fact, in \cite
{Banks:2002nv}\cite{Einhorn:2002nu}\cite{Kaloper:2002cs} it has been pointed
out that there could be tricky problems with field theories based on non
trivial vacua of this sort. In particular loop amplitudes are not
necessarily well defined. But as argued in \cite{Ulf:0210} (see also \cite
{Brandenberger:2002sr} for a similar point if view), the real world is not
expected to correspond to field theoretic $\alpha $-vacua up to infinite
energies. The best way to understand this is simply to view the system as an
excitation above the Bunch-Davies vacuum. What happens above the fundamental
scale we can only speculate about, but the hope is that the size, and as we
will see even the qualitative form, of the corrections (if there are any)
can be obtained without detailed knowledge of transplanckian physics.

The fact the effects are linear in $H/\Lambda $ is of crucial importance. If
the effects would come in at a higher power they will not be detectable
unless we invoke more exotic models of high energy physics involving large
extra dimensions. Discussions of such possibilities can be found in \cite
{Kaloper:2002uj}. What we will argue for here, is that rather conservative
models based on naturally occurring scales and expectations from string
theory give rise to effects that could be within reach.

At this point one should note, however, that even if we focus on linear
effects, there are nevertheless at least two different ways in which they
might be detectable. Either through corrections to the CMBR fluctuations, or
through high energy gamma rays. As discussed in \cite{Starobinsky:2002rp},
if one assumes that similar vacuum selection effects take place today as
during inflation -- after all the universe is still expanding -- this would
lead to the production of high energy radiation in \ the present universe.
It is argued in \cite{Starobinsky:2002rp} that this already puts stringent
limits on possible effects on the CMBR. While there is a certain tension
between these results we believe that the extrapolation over many orders of
magnitude from the inflationary period to the present day universe is
sufficiently uncertain that one can not make definite statements at the
present time. The most reasonable strategy seems to be to pursue both
approaches in looking for detectable effects.

To summarize, we believe that a detailed analysis of the possible effects on
the CMBR is of great interest in view of the upcoming MAP and Planck
satellites, and this is also the subject of the present paper.

The organization of the paper is as follows. In Section~\ref{sec:review} we
review the choice of initial conditions that we propose to investigate. In
Section~\ref{sec:look} we make some rough estimates of the kind of
signatures that one can expect from reasonable high energy physics. In
Section~\ref{sec:cmbr} we numerically analyze the corrections to the CMBR
spectrum and discuss whether the effects can be detected by MAP or Planck.
We end, finally, with some conclusions.

\section{Review of setup}

\label{sec:review}

\bigskip

In this section, following \cite{Danielsson:2002kx}, we will give a typical
example of the kind of corrections one might expect due to changes in the
low energy quantum state of the inflaton field due to transplanckian
effects. First we need to find out when to impose the initial conditions for
a mode with a given (constant) comoving momentum $k$. To do this it is
convenient to use conformal coordinates rather than the standard
Robertson-Walker coordinates where the inflating metric is given by

\begin{equation}
ds^{2}=dt^{2}-a\left( t\right) ^{2}dx^{2},
\end{equation}
with the scale factor given by $\ a\left( t\right) =e^{Ht}$. The conformal
coordinates are obtained by defining $\eta =-\frac{1}{aH}$ and the metric
takes the form 
\begin{equation}
ds^{2}=\frac{1}{H^{2}\eta ^{2}}\left( d\eta ^{2}-d\mathbf{x}^{2}\right) .
\label{confmet}
\end{equation}
We now note that the physical momentum $p$ and the comoving momentum $k$ are
related through 
\begin{equation}
k=ap=-\frac{p}{\eta H},
\end{equation}
and impose the initial conditions when $p=\Lambda $, where $\Lambda $ is the
energy scale important for the new physics. This scale could be the Planck
scale or the string scale. We find that the conformal time when the initial
condition is imposed to be 
\begin{equation}
\eta _{0}=-\frac{\Lambda }{Hk}.
\end{equation}
As we see, different modes will be created at different times, with a
smaller linear size of the mode (larger $k$) implying a later time.

To proceed we need the equation of motion for the scalar field (ignoring the
potential) 
\begin{equation}
\ddot{\phi}+3H\dot{\phi}-\nabla ^{2}\phi =0,
\end{equation}
which in terms of the conformal time $\eta $, and the rescaled field $\mu
=a\phi $, becomes 
\begin{equation}
\mu _{k}^{\prime \prime }+\left( k^{2}-\frac{a^{\prime \prime }}{a}\right)
\mu _{k}=0.  \label{modeeq}
\end{equation}
A nice discussion of the quantization of the system can be found in \cite
{Polarski:1995jg}. In terms of time dependent oscillators we can write

\begin{eqnarray}
\mu _{k}\left( \eta \right)  &=&\frac{1}{\sqrt{2k}}\left( a_{k}\left( \eta
\right) +a_{-k}^{\dagger }\left( \eta \right) \right)   \label{mupi} \\
\pi _{k}\left( \eta \right)  &=&\mu _{k}^{\prime }\left( \eta \right) +\frac{%
1}{\eta }\mu _{k}\left( \eta \right) =-i\sqrt{\frac{k}{2}}\left( a_{k}\left(
\eta \right) -a_{-k}^{\dagger }\left( \eta \right) \right) ,  \nonumber
\end{eqnarray}
which also can be expressed in terms of oscillators at a specific moment
using the Bogolubov transformation 
\begin{eqnarray}
a_{k}\left( \eta \right)  &=&u_{k}\left( \eta \right) a_{k}\left( \eta
_{0}\right) +v_{k}\left( \eta \right) a_{-k}^{\dagger }\left( \eta
_{0}\right)   \label{oscutv} \\
a_{-k}^{\dagger }\left( \eta \right)  &=&u_{k}^{\ast }\left( \eta \right)
a_{-k}^{\dagger }\left( \eta _{0}\right) +v_{k}^{\ast }\left( \eta \right)
a_{k}\left( \eta _{0}\right) ,  \nonumber
\end{eqnarray}
We find 
\begin{equation}
\mu _{k}\left( \eta \right) =f_{k}\left( \eta \right) a_{k}\left( \eta
_{0}\right) +f_{k}^{\ast }\left( \eta \right) a_{-k}^{\dagger }\left( \eta
_{0}\right) ,
\end{equation}
where

\bigskip 
\begin{equation}
f_{k}\left( \eta \right) =\frac{1}{\sqrt{2k}}\left( u_{k}\left( \eta \right)
+v_{k}^{\ast }\left( \eta \right) \right)
\end{equation}
is a solution of the mode equation (\ref{modeeq}). The solution can be
written as the linear combination

\begin{equation}
f_{k}=\frac{A_{k}}{\sqrt{2k}}e^{-ik\eta }\left( 1-\frac{i}{k\eta }\right) +%
\frac{B_{k}}{\sqrt{2k}}e^{ik\eta }\left( 1+\frac{i}{k\eta }\right) ,
\label{fk}
\end{equation}
where 
\begin{equation}
\left| A_{k}\right| ^{2}-\left| B_{k}\right| ^{2}=1.
\end{equation}

We are now in the position to start discussing the choice of vacuum. Without
knowledge of the high energy physics we can only list various possibilities
and investigate whether there is a typical size or signature of the new
effects. As an example, we will focus on a choice of vacuum determined by

\begin{equation}
a_{k}\left( \eta _{0}\right) \left| 0,\eta _{0}\right\rangle =0.
\end{equation}
This vacuum should be viewed as a typical representative of other vacua
besides the Bunch-Davies. It can be characterized as a vacuum corresponding
to a minimum uncertainty in the product of the field and its conjugate
momentum, \cite{Polarski:1995jg}, the vacuum with lowest energy (lower than
the Bunch-Davies) \cite{Starobinsky:2001kn}, or as the instantaneous
Minkowski vacuum. Therefore, it is as special as the Bunch Davies vacuum and
there is no a priori reason for planckian physics to prefer one over the
other.\footnote{%
In \cite{Niemeyer:2002kh} it is argued, based on adiabaticity, that even if
linear corrections are fine tuned away there will nevertheless remain
corrections of order \ $\left( H/\Lambda \right) ^{3}$ compared to the Bunch
Davies.}

We note that we in general have a class of different vacua depending on the
parameter $\eta _{0}$. \ At this initial time it follows from Eq. (\ref
{oscutv}) that $v_{k}\left( \eta _{0}\right) =0$, which, with the help of
Eq. (\ref{mupi}) and Eq. (\ref{fk}), will constrain $A_{k}$ and $B_{k}$ to
obey

\begin{equation}
B_{k}=\frac{ie^{-2ik\eta _{0}}}{2k\eta _{0}+i}A_{k}.
\end{equation}
Note that for $\eta _{0}\rightarrow -\infty $ one finds $A_{k}=1$ and $%
B_{k}=0$ which means that the Bunch-Davies vacuum is recovered. The reason
that the new vacuum is de Sitter invariant has to do with the way that the
initial conditions for a mode are imposed. The crucial point is that this is
done at a fixed scale, not a fixed time, and as a consequence physics will
actually be independent of time (up to changes in the inflationary
cosmological constant).

We are now in a position to calculate the expected fluctuation power
spectrum:

\begin{equation}
P(k)=\left( \frac{H}{\stackrel{\cdot }{\phi }}\right) ^{2}\left\langle
\left| \phi _{k}\left( \eta \right) \right| ^{2}\right\rangle =\left( \frac{H%
}{\stackrel{\cdot }{\phi }}\right) ^{2}\frac{1}{a^{2}}\left\langle \left|
\mu _{k}\left( \eta \right) \right| ^{2}\right\rangle =\left( \frac{H}{%
\stackrel{\cdot }{\phi }}\right) ^{2}\left( \frac{H}{2\pi }\right)
^{2}\left( 1-\frac{H}{\Lambda }\sin \left( \frac{2\Lambda }{H}\right)
\right) .  \label{eq:main}
\end{equation}
This result should be viewed as a typical example of what to be expected
from transplanckian physics if we allow for effects which at low energies
reduce to changes compared to the Bunch-Davies case. We note that the size
of the correction is linear in $H/\Lambda $, and that a Hubble constant that
varies during inflation give rise to a modulation of the spectrum. As argued
in \ \cite{Danielsson:2002kx}, this is presumably a quite generic effect
that is present regardless of the details of the transplanckian physics.
(See also \cite{Niemeyer:2002kh} for a discussion about this). After being
created at the fundamental scale the modes oscillate a number of times
before they freeze. The number of oscillations depend on the size of the
inflationary horizon and therefore changes when $H$ changes.

\section{What to look for}

\label{sec:look}

As discussed in the previous section, a Hubble constant that does not vary
during inflation would just imply a small change in the overall amplitude of
the fluctuation spectrum which would not constitute a useful signal.
Luckily, since the Hubble constant is expected to vary, the situation is
much more interesting.\footnote{%
In \cite{Easther:2002xe} the case of slow roll was also studied. For our
purposes, however, a small $\varepsilon $ expansion around the de Sitter
case is adequate.}

We will discuss what happens using slow roll parameters, see, e.g., \cite
{liddle}, where in particular, 
\begin{equation}
\varepsilon =\frac{M_{pl}^{2}}{2}\left( \frac{V^{\prime }}{V}\right) ^{2},
\end{equation}
with $M_{pl}=1/\sqrt{8\pi G}\sim 2\cdot 10^{18}GeV$ as the (reduced) Planck
mass. It is not difficult to show (using that $H$ is to be evaluated when a
given mode crosses the horizon, $k=aH$) that 
\begin{equation}
\frac{dH}{dk}=-\frac{\varepsilon H}{k},
\end{equation}
which gives 
\begin{equation}
H\sim k^{-\varepsilon }.
\end{equation}
The $k$ dependence of $H$ will translate into a modulation of $P(k)$, with a
periodicity given by 
\begin{equation}
\frac{\Delta k}{k}\sim \frac{\pi H}{\varepsilon \Lambda }.
\end{equation}
The overall amplitude of the CMBR spectrum (at the largest scales, still
outside the causal horizon at the time of photon decoupling) is given by 
\begin{equation}
\left( \frac{H}{\stackrel{\cdot }{\phi }}\right) ^{2}\left( \frac{H}{2\pi }%
\right) ^{2}=\frac{1}{24\pi ^{2}M_{pl}^{4}}\frac{V}{\varepsilon },
\end{equation}
which is restricted through measurements according to 
\begin{equation}
\frac{V^{1/4}}{\varepsilon ^{1/4}}\sim 0.027M_{pl}\equiv \beta M_{pl}.
\end{equation}
Using the Friedmann equation one finds 
\begin{equation}
H^{2}=\frac{V}{3M_{pl}^{2}}\sim \frac{\beta ^{4}}{3}M_{pl}^{2}\varepsilon ,
\end{equation}
from which it follows that 
\begin{equation}
\frac{H}{M_{pl}}\sim \frac{\beta ^{2}\varepsilon ^{1/2}}{\sqrt{3}}\sim
4\cdot 10^{-4}\sqrt{\varepsilon }.
\end{equation}
We now put the scale where initial conditions are imposed to be 
\begin{equation}
\Lambda =\gamma M_{pl},
\end{equation}
which implies that 
\begin{equation}
\frac{\Delta k}{k}\sim \frac{\pi H}{\varepsilon \gamma M_{pl}}\sim \frac{\pi
\beta ^{2}}{\sqrt{3}\gamma \sqrt{\varepsilon }}\sim 1.3\cdot 10^{-3}\frac{1}{%
\gamma \sqrt{\varepsilon }},
\end{equation}
and 
\begin{equation}
\xi \equiv \frac{H}{\Lambda }\sim 4\cdot 10^{-4}\frac{\sqrt{\varepsilon }}{%
\gamma }.  \label{eq:effect}
\end{equation}
To be more specific, we will now consider a realistic example. In the Ho\v{r}%
ava-Witten model \cite{Horava:1995qa}\cite{Horava:1996ma}, unification
occurs roughly at the same scale as a fifth dimension becomes visible and
also comparable to the string scale and the higher dimensional Planck scale.
For a discussion and references see, e.g., \cite{Polchinski:rr} or \cite
{Kaloper:2002uj}. As a rough estimate we therefore put $\Lambda =2\cdot
10^{16}$ GeV. This is a rather reasonable possibility within the framework
of the heterotic string and corresponds to $\gamma =0.01$. The Hubble
constant during inflation can not be much larger than $H=7\cdot 10^{13}$
GeV, corresponding to $\varepsilon =0.01$. Using these values we find 
\begin{eqnarray*}
\xi  &\sim &0.004 \\
\frac{\Delta k}{k} &=&\Delta \ln k\sim 1.
\end{eqnarray*}
This means one oscillation per logarithmic interval in $k$, which should be
visible in high-precision CMBR observation experiments. In the next section
we will consider these predictions in more detail.

\bigskip

\section{Predictions for CMBR measurements}

\label{sec:cmbr}

\bigskip

Density perturbations generated during the inflationary phase will imprint
corresponding temperature fluctuations in the cosmic microwave background
radiation (CMBR). These temperature fluctuations on the sky which are
gaussian due to their origin in oscillations of the inflation field can be
characterized by their angular power spectrum. (For a review, see e.g.~\cite
{kosowsky}). Expanding the temperature fluctuation in spherical harmonics
(excluding the dipole, which is indistinguishable from effects caused by our
motion with respect to the cosmic rest frame), 
\begin{equation}
{\frac{\Delta T}{T}}\left(\theta,\phi\right)=\sum_{\ell=2}^{\infty}
\sum_{m=-\ell}^{m=\ell}a_{\ell m}Y_{\ell m}\left(\theta,\phi\right)
\end{equation}
the angular power spectrum $C_\ell$ is defined by 
\begin{equation}
\langle a^*_{\ell m}a_{\ell^{\prime}m^{\prime}} \rangle = C_\ell
\delta_{\ell\ell^{\prime}}\delta_{mm^{\prime}},
\end{equation}
where the Kronecker deltas appear thanks to statistical isotropy. Since we
only have one sky to observe, an unbiased estimator for the angular power
spectrum, assuming full sky coverage, is given by 
\begin{equation}
C_\ell = {\frac{1}{2\ell +1}}\sum_{m=-\ell}^{m=\ell}a^*_{\ell m}a_{\ell m}.
\label{eq:cl}
\end{equation}
Well-known physics in the radiation dominated era and at the epoch near
photon decoupling will modify the ``primordial'' power spectrum in momentum
space $P(k)$ through a transfer function to the measured angular power
spectrum $C_\ell$. In particular, there will be a series of ``acoustic
peaks'', of which the location of the first one is an excellent indicator of
the overall geometry of the Universe. In the past few years, a number of
balloon-borne and terrestrial experiments have mapped out an ever larger
region in $\ell$-space. In particular, the location of the first peak has
been determined by the experiments BOOMERANG \cite{boomerang}, MAXIMA \cite
{maxima} and DASI \cite{dasi} to give clear evidence for a nearly flat
Universe. This has very recently been confirmed with even higher accuracy by
Archeops \cite{archeops}, and consistency of the CMBR data with the standard
scenario, including the inflationary mechanism, has been further
strengthened by the recent detection of a polarization signal by DASI \cite
{dasi2}. The satellite experiment MAP \cite{map} is expected to give results
soon, and a very ambitious mission, the Planck satellite, is being built for
a scheduled launch in 2007 \cite{planck}.

It is expected that the measurements of the angular power spectrum of the
CMBR encoded in the measured set of $C_\ell$ from satellites like Planck,
although superior to all present measurements, at most will reach an
accuracy of $10^{-2}$ for a particular value of $\ell$. This is essentially
set by cosmic variance, which is the fact that we only can sample one sky
from one particular point of observation (see Eq.~\ref{eq:cl}). This makes
it impossible \cite{knox} to reach a relative accuracy better than about $1/%
\sqrt{(2\ell+1)f_{\mathrm{sky}}}$, where $f_{\mathrm{sky}}$ is the fraction
of the sky covered by the experiment (which, for satellite experiments, will
be close to unity -- around 0.8 for Planck). This stochastic limitation of
the attainable accuracy is an order of magnitude above the estimate of the
transplanckian signal of the previous section. However, one has to take into
account the fact that future CMBR satellites will measure thousands of
independent data points corresponding to different $\ell$ values.

The transplanckian effects, regardless of their precise nature, have a
rather generic signature in form of their modulation of the spectrum. This
means that these effects will not likely be degenerate with a change in the
power spectrum caused by uncertainties in other parameters such as the slope
of the primordial spectrum $P(k)$ (which gives a nearly monotonous change
with $\ell $ of the angular power spectrum $C_{\ell }$) or the baryon
fraction of the matter density (which also causes oscillatory behavior in
the angular power spectrum, but with the modulation closely tied to the
location of the acoustic peaks). If it had just been an overall shift or
tilt of the amplitude it would not have been possible to measure the effect
even if it had been considerably larger than the percentage level. The shift
would just have gone into a changed value or slope of, e.g., $H$. With a
definite signature we can use several measurement points throughout the
spectrum, as we now discuss. We will see that Planck might be able to detect
transplanckian effects at the $10^{-3}$ level, which would put the Ho\v{r}%
ava-Witten model within range, or at least tantalizingly close. (To make a
more definitive statement, a more careful analysis of covariance of the
transplanckian signature with a large number of other cosmological
parameters will be needed, something we leave for future work.)

Let us now proceed with a somewhat more careful analysis. We see from Eq.~(%
\ref{eq:effect}) that, for fixed $\varepsilon $, it is reasonable to expect
a range of values of $\gamma$ within the reach of Planck (or even MAP). Too
large $\gamma $ and the amplitude will be too small. Too small $\gamma $ and
the period will be too long, but in an intermediate region the effects
should be observable.

In our result for the transplanckian power spectrum, Eq.~(\ref{eq:main}) 
\begin{equation}
P(k)=\left( \frac{H}{\stackrel{\cdot }{\phi }}\right) ^{2}\left( \frac{H}{%
2\pi }\right) ^{2}\left( 1-\frac{H}{\Lambda }\sin \left( \frac{2\Lambda }{H}%
\right) \right) ,
\end{equation}
we can for simplicity assume that the product of the first and second
factors gives a scale-invariant spectrum, i.e., 
\begin{equation}
\left( \frac{H}{\stackrel{\cdot }{\phi }}\right) ^{2}\left( \frac{H}{2\pi }%
\right) ^{2}=\mathrm{const.}
\end{equation}
(This assumption is not crucial, but can be arranged by a suitable choice of
the inflaton potential parameters defining the slow-roll phase of
inflation.) We can then instead of the parameter pair $\varepsilon $ and $%
\gamma $ choose the pair $\varepsilon $ and $\xi ={\frac{H_{n}}{\Lambda }}$,
with $H_{n}$ the Hubble parameter evaluated when some particular scale $k_{n}
$ leaves the horizon (in practice, we will choose this scale to correspond
to the largest angular scales measurable in the CMBR). This gives 
\[
{\frac{H}{\Lambda }}=\xi \left( {\frac{k}{k_{n}}}\right) ^{-\varepsilon },
\]
with $\xi \sim 4\cdot 10^{-4}{\frac{\sqrt{\varepsilon }}{\gamma }}$, which
enables us to parametrize the power spectrum in terms of the small parameter 
$\xi $ as 
\begin{equation}
P(\varepsilon ,\xi ;k)=P_{0}(k)\left( 1-\xi \left( {\frac{k}{k_{n}}}\right)
^{-\varepsilon }\sin \left[ {\frac{2}{\xi }}\left( {\frac{k}{k_{n}}}\right)
^{\varepsilon }\right] \right) .
\end{equation}
Here $P_{0}(k)$ is a scale-invariant spectrum which we can use for
comparison in the numerical work. The advantage of using $\xi $ instead of $%
\gamma $ is that it is a small parameter ($\xi \sim 0.004$ in the Ho\v{r}%
ava-Witten case) which can be safely extrapolated to zero -- the
transplanckian effects will have an unobservably small amplitude in this
limit. The possible variation of $\varepsilon $ is limited by the overall
normalization of the observed temperature fluctuations, so effectively one
can choose to regard the transplanckian effects as being a one-parameter
family of modulating functions, with amplitude determined by the value of $%
\xi $. Since the parameter $\gamma $ has a simpler physical interpretation,
we will however also use that one in the discussion of our results.

\begin{figure}[!htb]
\begin{center}
\epsfig{file=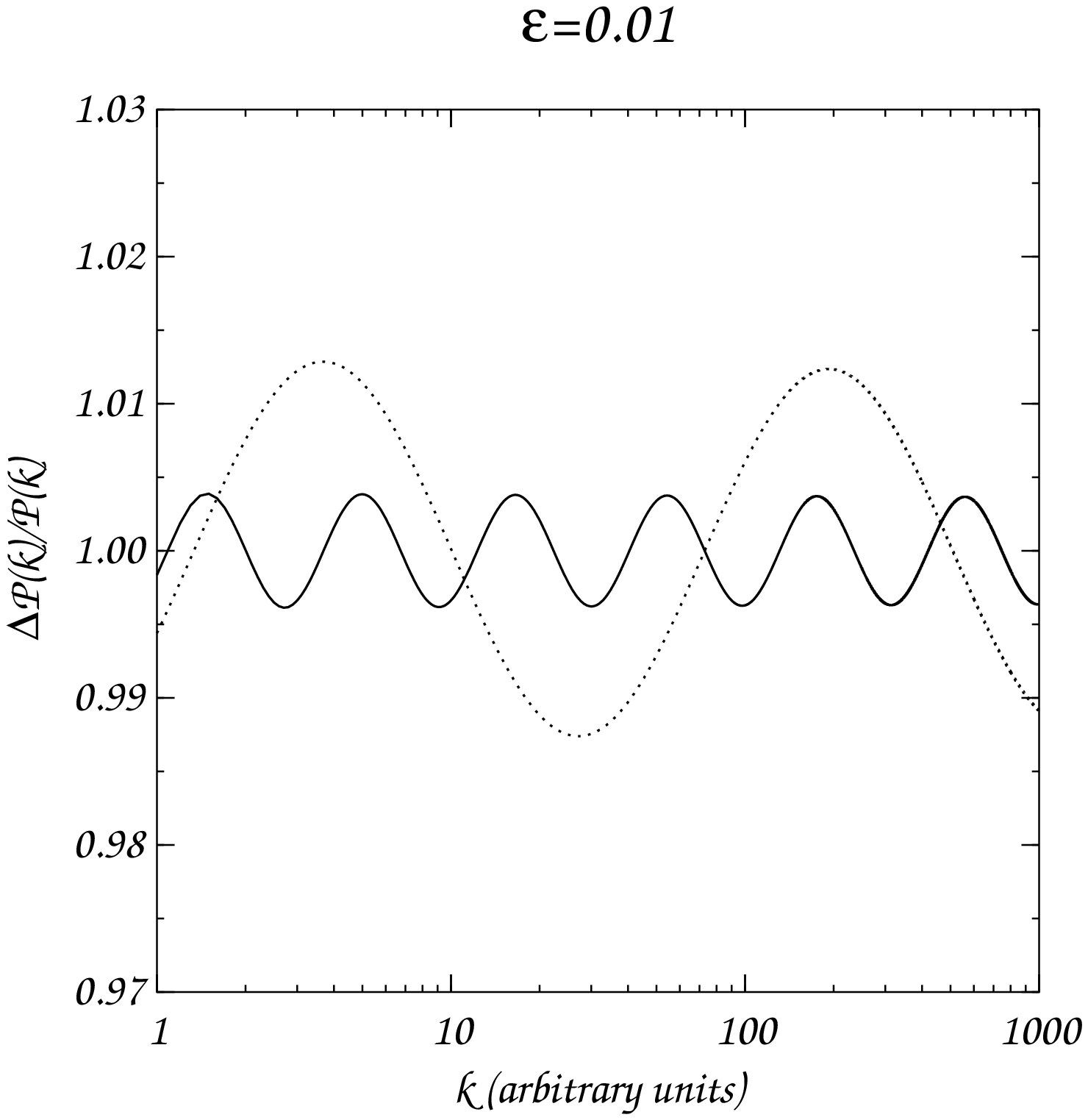,width=0.45\textwidth}%
\epsfig{file=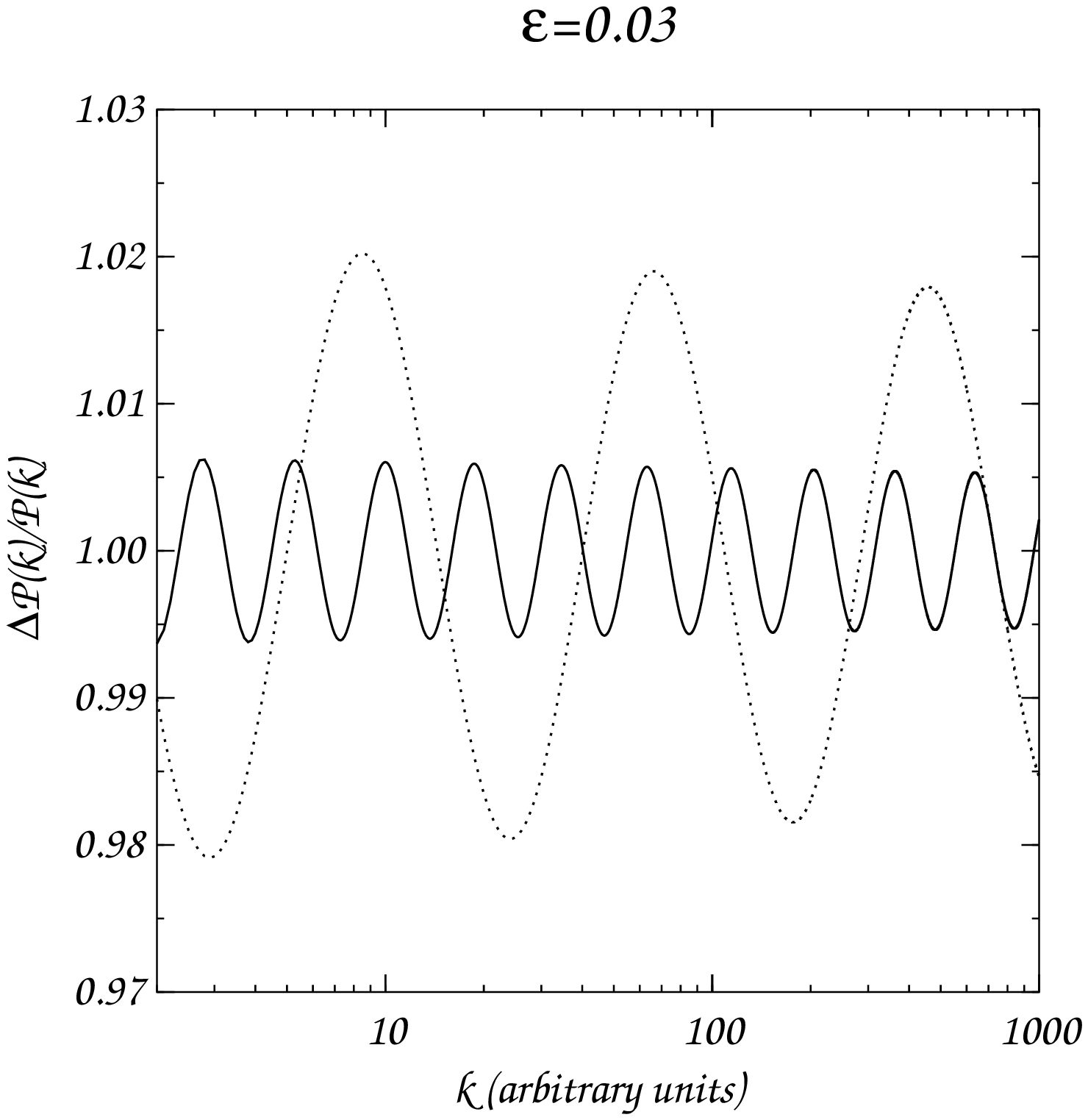,
width=0.45\textwidth}
\end{center}
\caption{(a) The modulation of the power spectrum of primordial density
fluctuations predicted in the transplanckian model (Eq.~(\ref{eq:main}))
with the parameter $\protect\varepsilon=0.01$ and $\protect\gamma = 0.01$
(solid line) and $\protect\gamma = 0.003$ (dotted line). (b) Same as in (a)
but with $\protect\varepsilon = 0.03$. }
\label{fig:fig1}
\end{figure}

In Fig.~\ref{fig:fig1} (a) and (b) we show examples of power spectra $P(k)$
for various values of $\gamma$, (and therefore of $\xi$) for $\varepsilon=$
0.01 and 0.03, respectively.

\begin{figure}[!htb]
\begin{center}
\epsfig{file=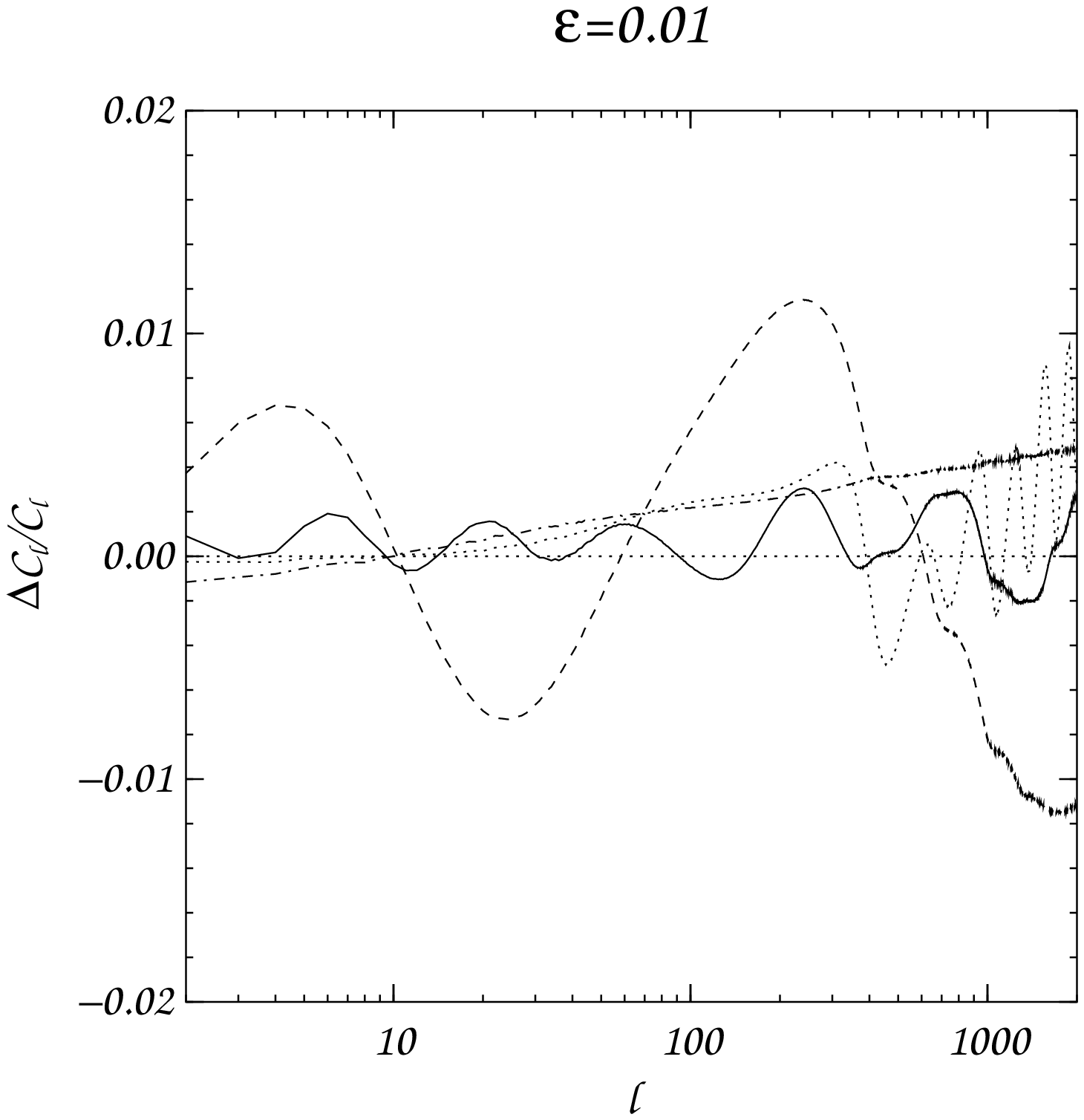,width=0.45\textwidth}%
\epsfig{file=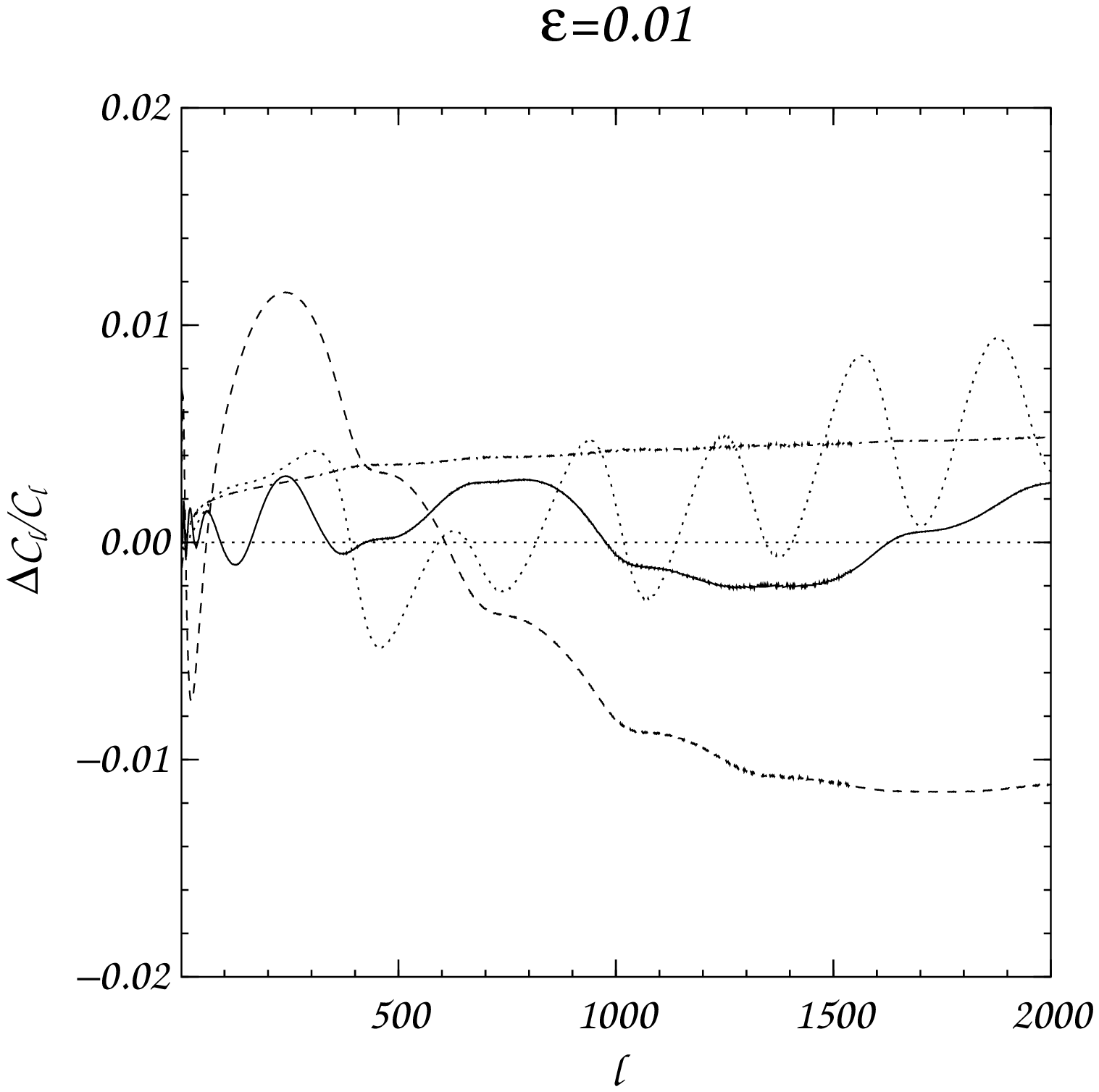,
width=0.45\textwidth}
\end{center}
\caption{(a) The relative change of the CMBR angular power spectrum $\Delta
C_\ell/C_\ell$ caused by transplanckian modulation of the primordial power
spectrum with $\protect\varepsilon =0.01$ and $\protect\gamma = 0.01$ (solid
line) and $\protect\gamma = 0.003$ (dashed line). For comparison, the
corresponding change caused by a change of the scale invariant spectrum $n=1$
to a slightly bluer one with $n=1.001$ is shown (the almost straight,
monotonous dash-dotted line). Also shown is the effect of increasing the
baryon fraction $\Omega_b$ by one percent, from 0.045 to 0.04545 (dotted
line). The scale in $\ell$ is logarithmic to facilitate comparison with the
modulation of the primordial spectrum shown in Fig.~\ref{fig:fig1}. The
small irregularities of the curves is caused by numerical noise in the
computations. (b) Same as (a), but with linear scale in $\ell$.}
\label{fig:fig2}
\end{figure}

To connect to observable quantities we have to compute the angular power
spectrum $C_\ell$ for each set of parameters. This is a complicated task
which has to be performed numerically. Fortunately, the form of the
transplanckian modulation makes it easy to modify existing computer codes to
incorporate the effects. We will use the well-known $\mathtt{cmbfast}$ code,
which is publicly available \cite{cmbfast}. As a first illustration of the
effects one may expect in the angular power spectrum of the CMBR, we show in
Fig.~\ref{fig:fig2} how the transplanckian effects using the Ho\v
rava-Witten parameters ($\varepsilon=0.01$, $\gamma=0.01$ meaning $\xi=0.004$%
) will appear, as well as a slightly lower $\gamma$, 0.003, ($\xi=0.013$)
which gives a larger effect. Shown is the fractional change $\Delta
C_\ell/C_\ell$ as a function of $\ell$, where we always will compare with an
``unperturbed'' model with scale-invariant power spectrum, present Hubble
constant 65 km/s Mpc$^{-1}$, matter density parameter $\Omega_M = 0.3$ of
which baryons contribute 0.045 and Cold Dark Matter 0.255, present
cosmological constant contribution $\Omega_\Lambda=0.7$, and no
reionization. Since we only will investigate small changes around a given
model, the exact choice of parameters for the reference model is not
essential. Also, for the same reason we consider it safe to display the
trends caused by the transplanckian modulation although they often are below
the nominal accuracy of the $\mathtt{cmbfast}$ program, which is around a
percent or so. (More accurate codes will no doubt appear well in time before
the launch of Planck.) The scale for $\ell$ in Fig.~\ref{fig:fig2} (a) is
logarithmic to show that much of the modulation apparent in Fig.~\ref
{fig:fig1} has survived being submitted by the transfer function to the CMBR
angular power spectrum.

For comparison, we also show in the same Figure the changes caused by a tilt
of the power spectrum by 0.001, and a change in the baryon density (or $%
\Omega_b h^2$, which is a more relevant physical quantity) by one percent
(keeping the total matter density unchanged). As can be seen, the effects
caused by the transplanckian modulation are visibly very different from the
other ones (we have verified that this is the case also for the other
``standard'' CMBR parameters). In Fig.~\ref{fig:fig2} (b) the same results
are shown using instead a linear scale for $\ell$.

We now want to get a quantitative measure of the potential observability of
the effects. The machinery of extracting cosmological parameters from the
CMBR is well developed by now (see, e.g., \cite
{hu,tegmark,kamionkowski,lineweaver,copeland,turner,steen}). Given a
``true'' underlying cosmology with a definite set of parameter values $s^0_i$%
, $i=1,\ldots N$, the probability to measure a given set of $C_\ell$
corresponding to fitted parameters $s_i$, which deviate from the true values
by $\delta s_i = s_i - s^0_i$, can be computed from the likelihood function 
\begin{equation}
\mathcal{L}=\mathcal{L}_m e^{-{\frac{1}{2}}\sum_{ij}F_{ij}\delta s_i \delta
s_j},
\end{equation}
where the ``Fisher matrix'' is given by 
\begin{equation}
F_{ij}=\sum_\ell{\frac{1}{\left(\Delta C^{\mathrm{obs}}_\ell\right)^2}}{%
\frac{\partial C_\ell}{\partial s_i}}{\frac{\partial C_\ell}{\partial s_j}}.
\end{equation}
Here the variance $\left(\Delta C^{\mathrm{obs}}_\ell\right)^2$ gets a
contribution both from the cosmic variance discussed above and from the
instrumental noise related to the angular resolution and sensitivity of the
experiment:

\begin{equation}
\left(\Delta C_\ell^{\mathrm{obs}}\right)^2\sim {\frac{2}{(2\ell +1)f_{%
\mathrm{sky}}}} \left(C_\ell + D_\ell\right)^2,
\end{equation}
where 
\begin{equation}
D_\ell= {\frac{1}{\sum_c w_ce^{-\ell(\ell+1)/\ell_c^2}}}.
\end{equation}
The experimental quantities $w_c$ and $\ell_c$ depend on the frequency
channel $c$ of the instrument. For Planck, we will use the first three HFI
channels with parameters as given in \cite{tegmark}.

As a first reality check of the observability of the transplanckian
modulation we assume that all other parameters are kept at the default
values, and we use hypothetical measurements from the Planck satellite to
determine with which significance we can say for a given value of $%
\varepsilon $ (which is more or less determined by the overall magnitude of
the fluctuations on the largest, COBE, scales) that $\xi $ is different from
zero. In this case the likelihood function collapses to 
\[
\mathcal{L}/\mathcal{L}_{m}=e^{-{\frac{1}{2}}\sum_{\ell }\left( {\frac{%
\Delta C_{\ell }^{\xi }}{\Delta C_{\ell }^{\mathrm{obs}}}}\right) ^{2}}, 
\]
where $\Delta C_{\ell }^{\xi }$ is the deviation from the reference model
caused by the modulation. For the Ho\v{r}ava-Witten values of the parameters
($\varepsilon =0.01$, $\xi =0.004$) we find that in this (idealized) case of
knowing all other parameters, the effect is right at the limit of being
observable with the likelihood of a chance occurrence of such a modulation
being around 17 \%. For a slightly larger value of $\xi $, which can be
obtained by a larger $\varepsilon $ and/or smaller $\gamma $, the
transplanckian modulation would be clearly visible. For instance, with the
same value of $\varepsilon $ and $\gamma =0.003$, the effect would have a
significance of around 10 standard deviations. Likewise, for $\varepsilon
=0.03$ and $\gamma =0.01$, the significance is around 3 standard deviations.
For the favorable case $\varepsilon =0.03$, $\gamma =0.003$ (or $\xi =0.02$,
still well within reasonable bounds of what one can expect for the unknown
transplanckian physics) the significance of the modulation would approach 16
standard deviations.

These estimates makes us believe that transplanckian effects, for $\xi$
values not too much smaller than 0.01, will turn out to be observable also
after a more complete analysis which includes a full Fisher matrix treatment
of covariance of parameters. In fact, by inspection of Fig.~\ref{fig:fig2}
we see that the effects for $\varepsilon = 0.01$ and $\gamma = 0.003$ ($%
\xi\sim 0.013$) are visibly larger than those obtained by changing the
baryon fraction by 1 \%, and the latter is definitely within reach of the
Planck satellite as shown in the analysis of \cite{tegmark}.

To make a first step towards a more complete analysis, we have computed the
2 by 2 Fisher matrix obtained by letting both $\xi$ and the only other
quantity which causes a modulation -- although of a very different type as
we have seen in Fig~\ref{fig:fig2} -- namely $\Omega_b$ vary. We confirm
that also in this case the one-sigma exclusion limit for $\xi$ is of the
order of 0.002.

For the case $\varepsilon = 0.03$, $\gamma = 0.003$ ($\xi=0.023$), the
effects could already start to be noticeable in the MAP experiment. Using
the parameters listed as MAP$^+$ in \cite{tegmark}, we obtain a calculated
significance between 3 and 4 standard deviations. To judge whether this is
observable in the real MAP experiment, a more realistic treatment of both
the experimental sensitivity and the effects of other cosmological
parameters would be called for. Doing the simple 2 by 2 Fisher matrix
analysis again, we find that the effect still is visible, but with only
around 2 sigma confidence. In any case, we find it intriguing that
transplanckian effects may be at the doorstep of current CMBR observations.

Thus, the CMBR may hold the clue to one of the most exciting problems of
today's theoretical physics, that of what happens to space-time for
distances smaller than the string or Planck lengths.

\section{Conclusions}

\bigskip

We have shown that in one of the simplest and in a sense most natural
scenarios which encodes our ignorance about physics at or near the Planck
scales, the CMBR appears as a very interesting source of information which
may give the first glimpses of how Nature works at its smallest -- and
largest -- scales. We believe that the results presented are very
encouraging. A careful study of present and future data should be
undertaken, looking for possible modulations. It is a new kind of ``exotic''
feature with distinct signature that should be experimentally investigated
irrespective of the state of the theoretical discussion which has not yet
reached a consensus at the present time. Even a negative result will put
meaningful restrictions on physics near the string or Planck scale, and
provide valuable guidance to the theoretical work.

\section*{Acknowledgments}

We would like to thank J.~Edsj\"{o}, A.~Lasenby, H.~Rubinstein and U.~Seljak
for useful discussions.
UD is a Royal Swedish Academy of Sciences Research Fellow supported by a
grant from the Knut and Alice Wallenberg Foundation. The work was also
supported by the Swedish Research Council (VR).


\bigskip

\begin{thebibliography}{99}
\bibitem{Brandenberger:1999sw}  R.~H.~Brandenberger, ``Inflationary
cosmology: Progress and problems,'' arXiv:hep-ph/9910410. 


\bibitem{Martin:2000xs}  J.~Martin and R.~H.~Brandenberger, ``The
trans-Planckian problem of inflationary cosmology,'' Phys.\ Rev.\ D \textbf{%
63}, 123501 (2001) [arXiv:hep-th/0005209]. 


\bibitem{Niemeyer:2000eh}  J.~C.~Niemeyer, ``Inflation with a high frequency
cutoff,'' Phys.\ Rev.\ D \textbf{63}, 123502 (2001)
[arXiv:astro-ph/0005533]. 


\bibitem{Brandenberger:2000wr}  R.~H.~Brandenberger and J.~Martin, ``The
robustness of inflation to changes in super-Planck-scale physics,'' Mod.\
Phys.\ Lett.\ A \textbf{16}, 999 (2001) [arXiv:astro-ph/0005432]. 


\bibitem{Kempf:2000ac}  A.~Kempf, ``Mode generating mechanism in inflation
with cutoff,'' Phys.\ Rev.\ D \textbf{63}, 083514 (2001)
[arXiv:astro-ph/0009209]. 


\bibitem{Chu:2000ww}  C.~S.~Chu, B.~R.~Greene and G.~Shiu, ``Remarks on
inflation and noncommutative geometry,'' Mod.\ Phys.\ Lett.\ A \textbf{16},
2231 (2001) [arXiv:hep-th/0011241]. 


\bibitem{Martin:2000bv}  J.~Martin and R.~H.~Brandenberger, ``A cosmological
window on trans-Planckian physics,'' arXiv:astro-ph/0012031. 

\bibitem{Mersini:2001su}  L.~Mersini, M.~Bastero-Gil and P.~Kanti, ``Relic
dark energy from trans-Planckian regime,'' Phys.\ Rev.\ D \textbf{64},
043508 (2001) [arXiv:hep-ph/0101210]. 

\bibitem{Niemeyer:2001qe}  J.~C.~Niemeyer and R.~Parentani,
``Trans-Planckian dispersion and scale-invariance of inflationary
perturbations,'' Phys.\ Rev.\ D \textbf{64}, 101301 (2001)
[arXiv:astro-ph/0101451]. 


\bibitem{Kempf:2001fa}  A.~Kempf and J.~C.~Niemeyer, ``Perturbation spectrum
in inflation with cutoff,'' Phys.\ Rev.\ D \textbf{64}, 103501 (2001)
[arXiv:astro-ph/0103225]. 

\bibitem{Starobinsky:2001kn}  A.~A.~Starobinsky, ``Robustness of the
inflationary perturbation spectrum to trans-Planckian physics,'' Pisma Zh.\
Eksp.\ Teor.\ Fiz.\ \textbf{73}, 415 (2001) [JETP Lett.\ \textbf{73}, 371
(2001)] [arXiv:astro-ph/0104043].

\bibitem{Easther:2001fi}  R.~Easther, B.~R.~Greene, W.~H.~Kinney and
G.~Shiu, ``Inflation as a probe of short distance physics,'' Phys.\ Rev.\ D 
\textbf{64}, 103502 (2001) [arXiv:hep-th/0104102]. 

\bibitem{Bastero-Gil:2001rv}  M.~Bastero-Gil and L.~Mersini, ``SN1A data and
CMB of Modified Curvature at Short and Large Distances,'' Phys.\ Rev.\ D 
\textbf{65} (2002) 023502 [arXiv:astro-ph/0107256]. 

\bibitem{Hui:2001ce}  L.~Hui and W.~H.~Kinney, ``Short distance physics and
the consistency relation for scalar and tensor fluctuations in the
inflationary universe,'' arXiv:astro-ph/0109107. 

\bibitem{Easther:2001fz}  R.~Easther, B.~R.~Greene, W.~H.~Kinney and
G.~Shiu, ``Imprints of short distance physics on inflationary cosmology,''
arXiv:hep-th/0110226. 

\bibitem{Bastero-Gil:2001nu}  M.~Bastero-Gil, P.~H.~Frampton and L.~Mersini,
``Modified dispersion relations from closed strings in toroidal cosmology,''
arXiv:hep-th/0110167. 

\bibitem{Brandenberger:2001ty}  R.~H.~Brandenberger, S.~E.~Joras and
J.~Martin, ``Trans-Planckian physics and the spectrum of fluctuations in a
bouncing universe,'' arXiv:hep-th/0112122. 

\bibitem{Martin:2002kt}  J.~Martin and R.~H.~Brandenberger, ``The
Corley-Jacobson dispersion relation and trans-Planckian inflation,''
arXiv:hep-th/0201189. 

\bibitem{Niemeyer:2002ze}  J.~C.~Niemeyer, ``Cosmological consequences of
short distance physics,'' arXiv:astro-ph/0201511. 

\bibitem{Lizzi:2002ib}  F.~Lizzi, G.~Mangano, G.~Miele and M.~Peloso,
``Cosmological perturbations and short distance physics from noncommutative
geometry,'' arXiv:hep-th/0203099. 

\bibitem{Shiu:2002kg}  G.~Shiu and I.~Wasserman, ``On the signature of short
distance scale in the cosmic microwave background,'' arXiv:hep-th/0203113.

\bibitem{Brandenberger:2002nq}  R.~Brandenberger and P.~M.~Ho,
``Noncommutative spacetime, stringy spacetime uncertainty principle, and
density fluctuations,'' arXiv:hep-th/0203119. 

\bibitem{Shankaranarayanan:2002ax}  S.~Shankaranarayanan, ``Is there an
imprint of Planck scale physics on inflationary cosmology?,''
arXiv:gr-qc/0203060. 
 
\bibitem{Kaloper:2002uj}  N.~Kaloper, M.~Kleban, A.~E.~Lawrence and
S.~Shenker, ``Signatures of short distance physics in the cosmic microwave
background,'' arXiv:hep-th/0201158. 


\bibitem{Brandenberger:2002hs}  R.~H.~Brandenberger and J.~Martin, ``On
signatures of short distance physics in the cosmic microwave background,''
arXiv:hep-th/0202142. 

\bibitem{Hassan:2002qk}  S.~F.~Hassan and M.~S.~Sloth, ``Trans-Planckian
effects in inflationary cosmology and the modified uncertainty principle,''
arXiv:hep-th/0204110.

\bibitem{Danielsson:2002kx}  U.~H.~Danielsson, ``A note on inflation and
transplanckian physics,'' Phys.\ Rev.\ D \textbf{66}, 023511 (2002)
[arXiv:hep-th/0203198]. 

\bibitem{Easther:2002xe}  R.~Easther, B.~R.~Greene, W.~H.~Kinney and
G.~Shiu, ``A generic estimate of trans-Planckian modifications to the
primordial power spectrum in inflation,'' arXiv:hep-th/0204129.

\bibitem{Danielsson:2002qh}  U.~H.~Danielsson, ``Inflation, holography and
the choice of vacuum in de Sitter space,'' JHEP \textbf{0207}, 040 (2002)
[arXiv:hep-th/0205227]. 

\bibitem{Niemeyer:2002kh}  J.~C.~Niemeyer, R.~Parentani and D.~Campo,
``Minimal modifications of the primordial power spectrum from an adiabatic
short distance cutoff,'' arXiv:hep-th/0206149. 

\bibitem{Goldstein:2002fc}  K.~Goldstein and D.~A.~Lowe, ``Initial state
effects on the cosmic microwave background and trans-planckian physics,''
arXiv:hep-th/0208167. 

\bibitem{Ulf:0210}  U.~H.~Danielsson, ``On the consistency of de Sitter
vacua,'' ~hep-th/0210058.

\bibitem{Brandenberger:2002sr}  R.~H.~Brandenberger, ``Trans-Planckian
physics and inflationary cosmology,'' arXiv:hep-th/0210186.

\bibitem{chernikov}  N. A. Chernikov and E. A. Tagirov, ``Quantum theory of
scalar field in de Sitter space-time,'' Ann. Inst. Henri Poincar\'{e}, vol.
IX, nr 2, (1968) 109.

\bibitem{Mottola:ar}  E.~Mottola, ``Particle Creation In De Sitter Space,''
Phys.\ Rev.\ D \textbf{31} (1985) 754.

\bibitem{Allen:ux}  B.~Allen, ``Vacuum States In De Sitter Space,'' Phys.\
Rev.\ D \textbf{32} (1985) 3136.

\bibitem{Floreanini:1986tq}  R.~Floreanini, C.~T.~Hill and R.~Jackiw,
``Functional Representation For The Isometries Of De Sitter Space,'' Annals
Phys.\ \textbf{175} (1987) 345.

\bibitem{Bousso:2001mw}  R.~Bousso, A.~Maloney and A.~Strominger,
``Conformal vacua and entropy in de Sitter space,'' arXiv:hep-th/0112218.

\bibitem{Spradlin:2001nb}  M.~Spradlin and A.~Volovich, ``Vacuum states and
the S-matrix in dS/CFT,'' arXiv:hep-th/0112223.

\bibitem{Banks:2002nv}  T.~Banks and L.~Mannelli, ``De Sitter vacua,
renormalization and locality,'' arXiv:hep-th/0209113. 

\bibitem{Einhorn:2002nu}  M.~B.~Einhorn and F.~Larsen, ``Interacting Quantum
Field Theory in de Sitter Vacua,'' arXiv:hep-th/0209159. 


\bibitem{Kaloper:2002cs}  N.~Kaloper, M.~Kleban, A.~Lawrence, S.~Shenker and
L.~Susskind, ``Initial conditions for inflation,'' arXiv:hep-th/0209231.


\bibitem{Starobinsky:2002rp}  A.~A.~Starobinsky and I.~I.~Tkachev,
``Trans-Planckian particle creation in cosmology and ultra-high energy
cosmic rays,'' arXiv:astro-ph/0207572. 

\bibitem{Polarski:1995jg}  D.~Polarski and A.~A.~Starobinsky,
``Semiclassicality and decoherence of cosmological perturbations,'' Class.\
Quant.\ Grav.\ \textbf{13}, 377 (1996) [arXiv:gr-qc/9504030].

\bibitem{liddle}  A. R. Liddle and D. H. Lyth, ``Cosmological inflation and
large-scale structure'', Cambridge University Press 2000.

\bibitem{Horava:1995qa}  P.~Horava and E.~Witten, ``Heterotic and type I
string dynamics from eleven dimensions,'' Nucl.\ Phys.\ B \textbf{460}
(1996) 506 [arXiv:hep-th/9510209]. 

\bibitem{Horava:1996ma}  P.~Horava and E.~Witten, ``Eleven-Dimensional
Supergravity on a Manifold with Boundary,'' Nucl.\ Phys.\ B \textbf{475}
(1996) 94 [arXiv:hep-th/9603142]. 


\bibitem{Polchinski:rr}  J.~Polchinski, ``String Theory. Vol. 2: Superstring
Theory And Beyond,'' \textit{Cambridge, UK: Univ. Pr. (1998) 531 p}.

\bibitem{kosowsky}  A.~Kosowsky, ``The cosmic microwave background,''
arXiv:astro-ph/0102402. 


\bibitem{boomerang}  C.~B.~Netterfield \textit{et al.} [Boomerang
Collaboration], ``A measurement by BOOMERANG of multiple peaks in the
angular power spectrum of the cosmic microwave background,'' Astrophys.\ J.\ 
\textbf{571}, 604 (2002) [arXiv:astro-ph/0104460]. 


\bibitem{maxima}  S.~Hanany \textit{et al.}, ``MAXIMA-1: A Measurement of
the Cosmic Microwave Background Anisotropy on angular scales of 10
arcminutes to 5 degrees,'' Astrophys.\ J.\ \textbf{545}, L5 (2000)
[arXiv:astro-ph/0005123]. 


\bibitem{dasi}  C.~Pryke, N.~W.~Halverson, E.~M.~Leitch, J.~Kovac,
J.~E.~Carlstrom, W.~L.~Holzapfel and M.~Dragovan, ``Cosmological Parameter
Extraction from the First Season of Observations with DASI,'' Astrophys.\
J.\ \textbf{568}, 46 (2002) [arXiv:astro-ph/0104490]. 


\bibitem{archeops}  A.~Benoit [the Archeops Collaboration], ``The Cosmic
Microwave Background Anisotropy Power Spectrum measured by Archeops,''
arXiv:astro-ph/0210305. 


\bibitem{dasi2}  E.~M.~Leitch \textit{et al.}, ``Measuring Polarization with
DASI,'' arXiv:astro-ph/0209476. 

\bibitem{map}  The NASA MAP mission, homepage http://map.gsfc.nasa.gov/

\bibitem{planck}  The ESA Planck mission, homepage
http://astro.estec.esa.nl/SA-general/Projects/Planck/

\bibitem{knox}  L.~Knox, ``Determination of inflationary observables by
cosmic microwave background anisotropy experiments,'' Phys.\ Rev.\ D \textbf{%
52}, 4307 (1995) [arXiv:astro-ph/9504054]. 


\bibitem{cmbfast}  U.~Seljak and M.~Zaldarriaga, ``A Line of Sight Approach
to Cosmic Microwave Background Anisotropies,'' Astrophys.\ J.\ \textbf{469},
437 (1996) [arXiv:astro-ph/9603033]. 


\bibitem{hu}  W.~Hu and N.~Sugiyama, ``Anisotropies in the cosmic microwave
background: An Analytic approach,'' Astrophys.\ J.\ \textbf{444}, 489
(1995). 


\bibitem{tegmark}  J.~R.~Bond, G.~Efstathiou and M.~Tegmark, ``Forecasting
Cosmic Parameter Errors from Microwave Background Anisotropy Experiments,''
Mon.\ Not.\ Roy.\ Ast.\ Soc.\ \textbf{291}, L33 (1997)
[arXiv:astro-ph/9702100]. 


\bibitem{kamionkowski}  G.~Jungman, M.~Kamionkowski, A.~Kosowsky and
D.~N.~Spergel, ``Cosmological parameter determination with microwave
background maps,'' Phys.\ Rev.\ D \textbf{54}, 1332 (1996)
[arXiv:astro-ph/9512139]. 


\bibitem{lineweaver}  C.~H.~Lineweaver, ``The Cosmic Microwave Background
and Observational Convergence in the $\Omega_m - \Omega_\lambda$ Plane,''
Astrophys.\ J.\ \textbf{505}, L69 (1998) [arXiv:astro-ph/9805326]. 


\bibitem{copeland}  E.~J.~Copeland, I.~J.~Grivell and A.~R.~Liddle,
``Cosmological parameter estimation and the spectral index from inflation,''
Mon.\ Not.\ Roy.\ Ast.\ Soc.\ \textbf{298}, 1233 (1998)
[arXiv:astro-ph/9712028]. 


\bibitem{turner}  T.~Souradeep, J.~R.~Bond, L.~Knox, G.~Efstathiou and
M.~S.~Turner, ``Prospects for measuring inflation parameters with the CMB,''
arXiv:astro-ph/9802262. 


\bibitem{steen}  S.~Hannestad, ``Reconstructing the inflationary power
spectrum from CMBR data,'' Phys.\ Rev.\ D \textbf{63}, 043009 (2001)
[arXiv:astro-ph/0009296]. 
\end{thebibliography}
\end{document}